\documentclass[12pt,a4paper,english,aps,manuscript,final,showpacs,pra,lengthcheck,superscriptaddress,groupaddress]{revtex4-1}
\usepackage{ae,aecompl}

\usepackage[T1]{fontenc}
\usepackage[latin9]{inputenc}
\usepackage{xcolor}
\usepackage{pdfcolmk}
\definecolor{document_fontcolor}{rgb}{0, 0, 0}
\color{document_fontcolor}
\usepackage{babel}
\usepackage{textcomp}
\usepackage{graphicx}
\PassOptionsToPackage{normalem}{ulem}
\usepackage{ulem}
\usepackage[unicode=true,pdfusetitle,
 bookmarks=true,bookmarksnumbered=false,bookmarksopen=false,
 breaklinks=false,pdfborder={0 0 1},backref=false,colorlinks=true]
 {hyperref}

\makeatletter

\pdfpageheight\paperheight
\pdfpagewidth\paperwidth

\providecolor{lyxadded}{rgb}{0,0,1}
\providecolor{lyxdeleted}{rgb}{1,0,0}
\newcommand{\lyxadded}[3]{{\texorpdfstring{\color{lyxadded}{}}{}#3}}

 \@ifundefined{textcolor}{}
 {%
   \definecolor{BLACK}{gray}{0}
   \definecolor{WHITE}{gray}{1}
   \definecolor{RED}{rgb}{1,0,0}
   \definecolor{GREEN}{rgb}{0,1,0}
   \definecolor{BLUE}{rgb}{0,0,1}
   \definecolor{CYAN}{cmyk}{1,0,0,0}
   \definecolor{MAGENTA}{cmyk}{0,1,0,0}
   \definecolor{YELLOW}{cmyk}{0,0,1,0}
 }

\makeatother

\begin{document}

\title{Robustness of dynamical decoupling sequences}

\author{Mustafa Ahmed Ali Ahmed}

\email{mustafa.ahmed@tu-dortmund.de}

\affiliation{Fakult\"at Physik, Technische Universit\"at Dortmund, Dortmund, Germany.}

\affiliation{Department of Physics, International University of Africa, Khartoum, Sudan }

\author{Gonzalo A. \'Alvarez}

\email{gonzalo.a.alvarez@weizmann.ac.il}

\affiliation{Fakult\"at Physik, Technische Universit\"at Dortmund, Dortmund, Germany.}

\affiliation{Department of Chemical Physics, Weizmann Institute of Science, Rehovot,
Israel}

\author{Dieter Suter}

\email{dieter.suter@tu-dortmund.de}

\affiliation{Fakult\"at Physik, Technische Universit\"at Dortmund, Dortmund, Germany.}
\begin{abstract}
Active protection of quantum states is an essential prerequisite for the
implementation of quantum computing. Dynamical decoupling (DD) is a promising
approach that applies sequences of control pulses to the system in order
to reduce the adverse effect of system-environment interactions\@. Since
every hardware device has finite precision, the errors of the DD control
pulses can themselves destroy the stored information rather than protect
it\@. We experimentally compare the performance of different DD sequences
in the presence of an environment that was chosen such that all relevant
DD sequences can equally suppress its effect on the system. Under these
conditions, the remaining decay of the qubits under DD allows us to compare
very precisely the robustness of the different DD sequences with respect
to imperfections of the control pulses.
\end{abstract}

\date{11/20/12}

\maketitle

\section{Introduction}

Quantum computers can execute certain tasks more efficiently than classical
computers by processing information according to the laws of quantum mechanics\@.
In analogy to a classical bit, which can assume the values $0$ or $1$,
quantum mechanical two-level systems like a spin-1/2 can be used as quantum
bits by identifying their eigenstates with these values, e.g. $\left|0\right\rangle $
for spin up and $\left|1\right\rangle $ for spin down\@. In quantum information
processing and quantum memory applications, it is very important to keep
the information isolated from the environment: uncontrolled interactions
with the environment tend to degrade the quantum information\@. This environment-induced
loss of quantum information is called `decoherence' \citep{Zurek2003,Nielsen2004,Kaye2007}\@.

If one is able to control the system in such a way to reduce the detrimental
effect of the system-environment (SE) interaction, one can preserve the
quantum state for a longer time\@. This way of fighting decoherence by
applying fast and strong pulses has been termed dynamical decoupling (DD)\citep{Viola1999,Khodjasteh2005,Uhrig2007,Gordon2008,Clausen2010,Yang2011}.
The main attraction of DD is that it requires few additional resources.
In contrast to quantum error correction \citep{Preskill1998,Cory1998,Knill1998},
e.g., it does not require additional qubits\@. DD can be traced back to
Hahn's `spin echo' experiment, where a refocusing pulse induces a time
reversal of the SE interaction of nuclear spins \citep{Hahn1950}\@. This
increases the decay time or decoherence time of the stored information
in the qubit. The technique has evolved significantly since then and its
efficiency was studied and demonstrated in many different systems \citep{Cywinski2008,Yang2008,Biercuk2009a,Du2009,Alvarez2010,Lange2010,Barthel2010,Pasini2010a,Ryan2010,Ajoy2011,Souza2011,Almog2011,RoyBardhan2012,Pan2011,Shukla2011,Bluhm2011,Naydenov2011}.

It has been shown that the type as well as the spectral density of the
SE interaction play a significant role for finding the optimal DD sequences
\citep{Kofman2001,Kofman2004,Zhang2007,Gordon2008,BhaktavatsalaRao2011,Bylander2011,Alvarez2011,Almog2011,Kotler2012,Bar-Gill2012}\@.
Moreover, unavoidable errors in the control pulses are also an important
source of decoherence \citep{Khodjasteh2007,Hodgson2010,Alvarez2010,Souza2011,Wang2011,Xiao2011,Khodjasteh2011,Peng2011,Wang2012,Souza2012b,Alvarez2012,Souza2012a,Souza2012}\@.
Thus optimal DD sequences must be able to reduce the effective SE interaction
while compensating the effects of non-ideal control fields \citep{Viola2003,Ryan2010,Souza2011,Souza2012a,Khodjasteh2011,Souza2012b,Cai2012,Alvarez2012,Souza2012}.
Reference \citep{Souza2012a} is a recent review of this subject.

In this article we compare the performance of different DD sequences in
a system where pulse errors are the dominant source of decoherence. To
the best of our knowledge, this situation has not yet been investigated:
all previous works were done in systems where the pulse errors only become
significant when most of the effects of the SE interaction have been eliminated.
To this end, we prepare a system where the spectral density of the SE coupling
has two main contributions. One source of noise is almost static and can
therefore be refocused by all tested DD sequences. The other contribution
is a rapidly fluctuating noise, whose correlation time is much shorter
than the time required for an inversion of the spins. This type of noise
cannot be refocused by any DD sequence. Therefore the main difference between
the performance of the DD sequences is their susceptibility to pulse errors. 

In a previous work \citep{Alvarez2010,Souza2011}, we compared the performance
of different DD sequences and in particular we found that the `Carr\textendash{}Purcell\textendash{}Meiboom\textendash{}Gill'
(CPMG) sequence \citep{Carr1954,Meiboom1958} performed particularly well
for specific initial states. In this case the decoherence time was one
order of magnitude longer than for robust sequences that reduced decoherence
symmetrically with respect to arbitrary initial states. We found that this
difference arose because the system qubit, a $^{13}$C spin, interacted
with neighboring $^{13}$C spins. While the CPMG sequence was able to reduce
the effect of $^{13}$C-$^{13}$C couplings, the robust sequences were
not \citep{Alvarez2011}\@. Since the tested DD sequences were not designed
for eliminating the effect of homonuclear couplings, the longer decoherence
times for CPMG applied to certain initial conditions cannot be taken as
a measure of its performance. In this work we therefore use a system that
does not exhibit homonuclear interactions. This allows us to show that
the robust sequences can also achieve the optimal decoherence time observed
under CPMG, and this performance is independent of the initial condition.

The article is organized as follows. In Sec. II we define the system, in
Sec. III we introduce the DD sequences to be compared and in Sec. IV we
compare their robustness against pulse errors effects. In Sec. V we give
a qualitative theoretical analysis based on average Hamiltonian theory
to explain the experimental results. In the last section we draw some conclusions.

\section{The system}

The experimental system is an ensemble of non-interacting spins 1/2. They
consist of the protons of a water sample to which we added 5~mg/100~ml
CuSO$_{4}$ to reduce the $T_{1}$ relaxation time to 287~ms. This results
in faster repetition times and shorter overall duration of the experiments.
The sample was placed in a static magnetic field along the $z$-direction
and its Hamiltonian is\lyxadded{mustafa}{Fri Mar 08 08:32:48 2013}{
\begin{equation}
\mathcal{H}_{s}=\omega_{s}S_{z},\label{eq:Hs}
\end{equation}
}where $\omega_{s}$ is the Zeeman frequency and $\widehat{S}_{z}$ is
the system spin operator along the $z$ axis\@. The inhomogeneities of
the static field correspond to a static perturbation, and molecular motion
makes this perturbation time-dependent on a time scale that is slow compared
with the delays between the DD pulses used in our experiments. This makes
it possible to refocus this perturbation very effectively. 

The second major source of noise is the fluctuating dipole-dipole interaction,
whose correlation time is the molecular reorientation time ($\approx$
35 fs) - much faster than any conceivable control fields for nuclear spins
and therefore not amenable to DD. On the other hand, these fluctuations
are so fast that their average effect on the system is relatively small
\citep{Bloembergen1948}.

\begin{figure}[h]
\centering{}\includegraphics[clip,width=1\columnwidth]{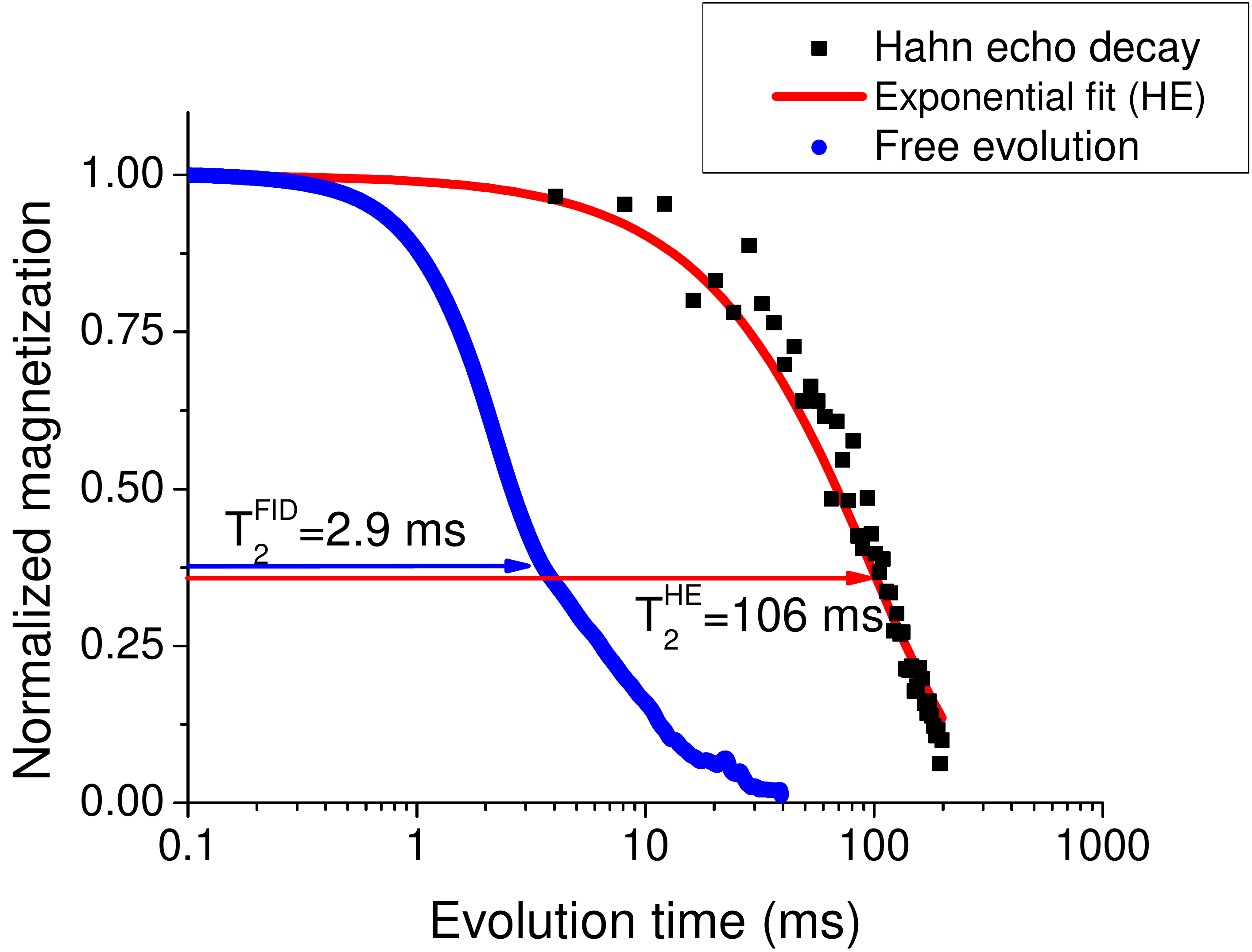}\caption{(Color online) Decay of the magnetization of the $^{1}$H spin system under
free evolution and in a Hahn spin-echo (HE) sequence.\label{1}}
\end{figure}

Experiments were performed on a home-built NMR spectrometer with a $^{1}$H
resonance frequency of 360~MHz. The radio frequency field strength was
$2\pi\cdot13.3$~kHz, which corresponds to a $\pi-$pulse duration of
37.5~\textmu{}s. An initial state $\propto I_{x}$ or $I_{y}$ was prepared
by rotating the $I_{z}$ equilibrium state with a resonant $\pi/2$ pulse.
The free evolution decay of the transversal magnetization of our system
has a decay time of 2.9~ms (free induction decay). A simple Hahn-echo
sequence \citep{Hahn1950} increases this time to 106 ms, as shown in Fig.
\ref{1}\@.

\section{Dynamical decoupling sequences}

DD sequences consist of repetitive trains of $\pi$-pulses. The delays
between the pulses and their phases are important parameters for improving
the performance of the DD sequences \citep{Carr1954,Meiboom1958,Maudsley1986,Gullion1990,Viola1999,Viola2003,Khodjasteh2005,Uhrig2007}.
In particular the relative phases, which correspond to the directions of
the rotation axes, are important for making the sequences robust against
pulse imperfections and unwanted environmental interactions \citep{Ryan2010,Alvarez2010,Souza2011}. 

\begin{figure}[h]
\begin{centering}
\includegraphics[bb=20bp 100bp 590bp 820bp,width=0.95\columnwidth]{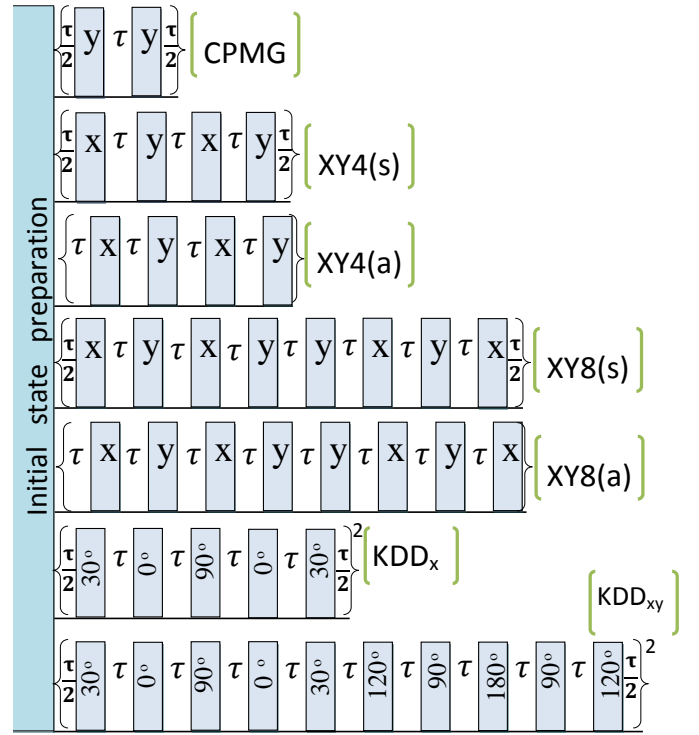}
\par\end{centering}

\caption{(Color online) Dynamical decoupling pulse sequences tested in this work.\label{2-2}}
\end{figure}

Fig. \ref{2-2} gives an overview of the sequences that we examined for
this work. It shows a single cycle for each sequence, which is repeated
as often as required. $\tau$ is the delay between the pulses\@. The Carr
- Purcell (CP) sequence \citep{Carr1954} and the version by Meiboom and
Gill \citep{Meiboom1958}, known as CPMG, use the same sequence of refocusing
pulses; they only differ with respect to the state to which they are applied.
In the case of the CP sequence, the initial state is perpendicular to the
rotation axis of the inversion pulses, in the CPMG version, it is parallel.
Errors in the flip angles destroy the perpendicular component, but they
leave the longitudinal component unscathed \citep{Meiboom1958,Alvarez2010}\@.
The sequence XY4 was introduced by Maudsley \citep{Maudsley1986} and it
reduces the effect of pulse imperfections for arbitrary initial states
\citep{Maudsley1986,Gullion1990,Alvarez2010}. It consists of four pulses
with phases $x-y-x-y$ (Fig. \ref{2-2})\@. An asymmetric version of the
XY-4 sequence was introduced by Viola et al. \citep{Viola1999}, which
we designate XY4(a). The XY8-sequences are symmetrized versions of the
XY4 sequences \citep{Gullion1990,Souza2012b}. Two DD sequences that are
particularly robust against flip-angle and resonance offset errors are
the KDD$_{x}$ and KDD$_{xy}$ sequences \citep{Souza2011,Alvarez2012,Souza2012a}.
They were designed by combining the rotation pattern of the XY4 sequence
with that of a robust composite pulse \citep{tycko:2775}.

In earlier works using these sequences, the conditions were chosen such
that the dominant perturbation was the environmental noise \citep{Alvarez2010,Souza2011,Peng2011,Souza2012b,Alvarez2012}.
In this work, we focus on a system that allow us to make a comparison between
these sequences in a regime where all sequences perform equally well at
eliminating the environmental noise and any differences in their performance
can be attributed directly to their robustness, i.e. to their efficiency
in suppressing the effect of pulse imperfections.

\section{Robustness comparison}

\begin{figure}[h]
\begin{centering}
\includegraphics[clip,width=1\columnwidth]{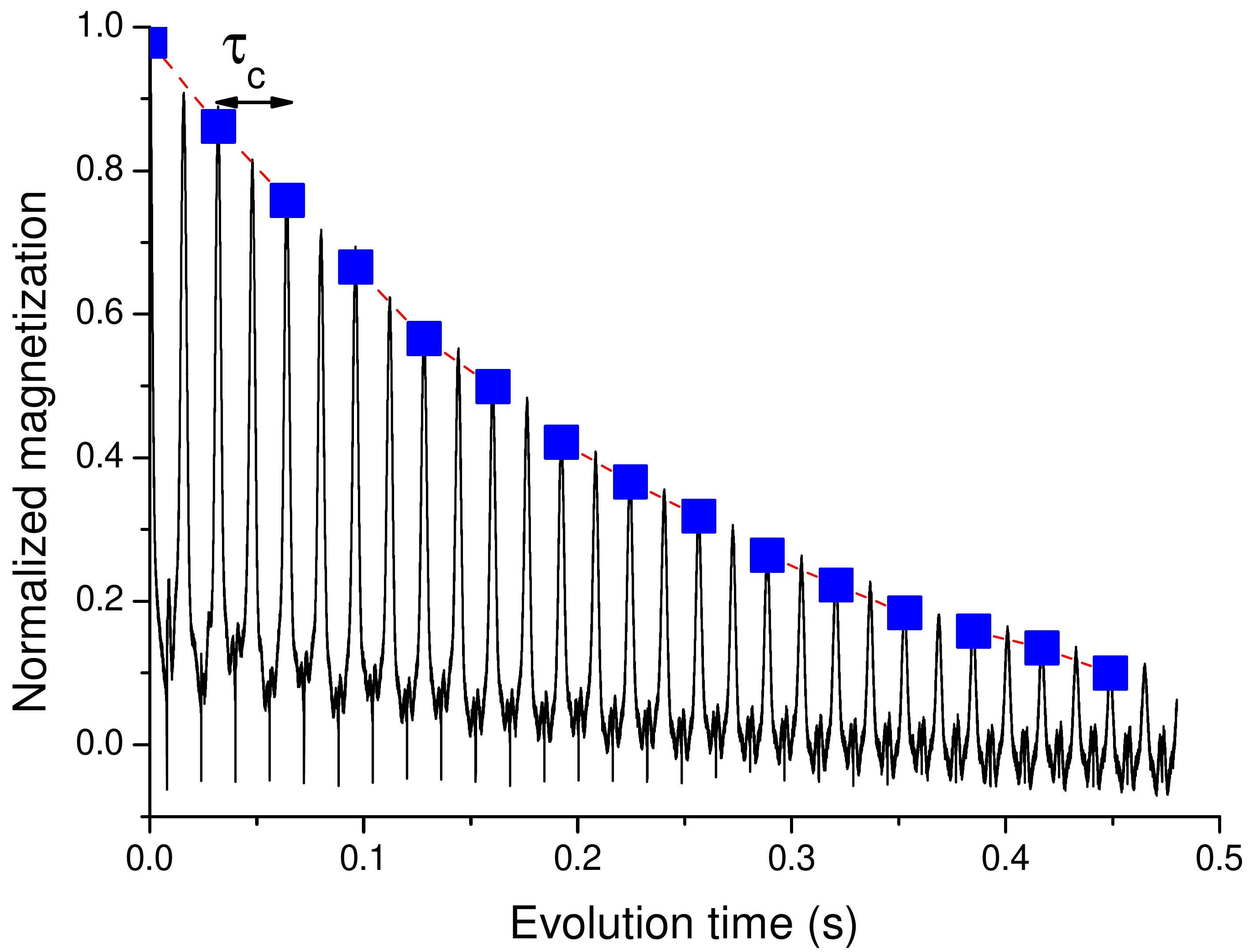}
\par\end{centering}

\caption{(Color online) Time evolution of the spin-system magnetization under the
application of a CPMG sequence. The black solid line shows the evolution
of the magnetization and the blue squares mark the echo amplitude at the
end of a CPMG cycle. We use the echo maxima for measuring the CPMG decay
time. The cycle time was $\tau_{c}=$ 32 ms.\label{3}}
\end{figure}

To compare the sensitivity of the sequences to pulse imperfections, we
prepared two orthogonal initial states $I_{x}$ and $I_{y}$ and then measured
their decay as a function of time under the application of the different
DD sequences described in the previous section. Figure \ref{3} shows the
echo train of a CPMG sequence\@. From this data, we extracted the signal
($I_{x}$ magnetization in this example) at the end of each DD cycle (marked
by blue squares in the figure). The decay of the echoes was mostly exponential,
with some exceptions discussed below.

The extracted echoes (blue squares in Fig.\ \ref{3}) were fitted with
an exponential function to obtain the decay time of the magnetization\@.
Experiments were repeated and the decay times plotted as a function of
the delay between pulses\@. 

\begin{figure}[h]
\centering{}\includegraphics[scale=0.3]{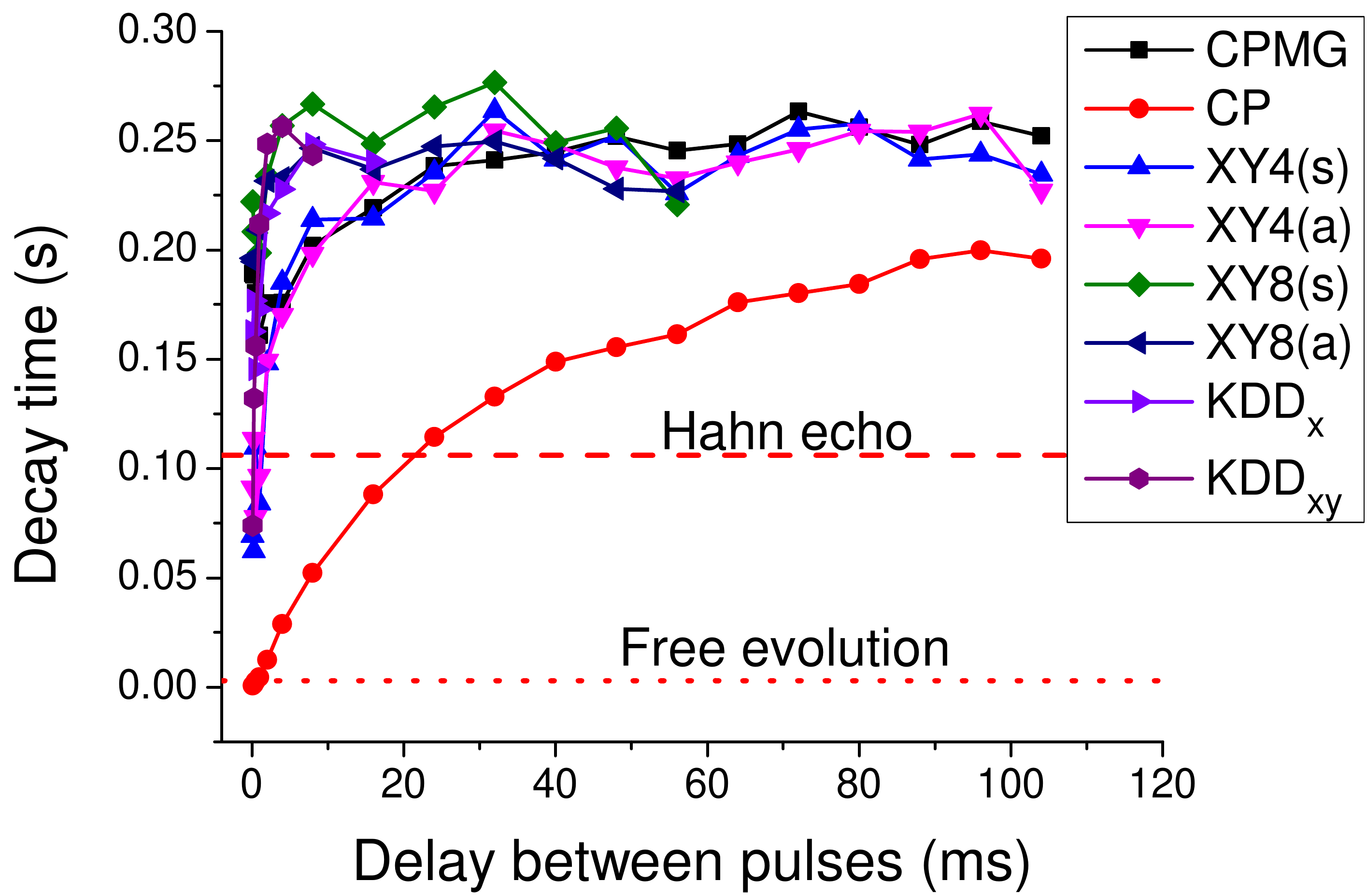}\caption{(Color online) Average decay times as a function of the delay $\tau$ between
pulses for different DD sequences \label{4}.}
\end{figure}

Figure \ref{4} compares the decay times of different DD sequences as a
function of the delay between pulses. For the CPMG sequence, we present
the decay of the $I_{x}$ and $I_{y}$ magnetization separately, marked
as CP and CPMG, respectively. For the other sequences, whose performance
is quite symmetric with respect to the initial condition, we present the
decay times averaged over the two initial conditions. For long delays between
the pulses, the observed decay times reach a limiting value of $\approx276$
ms, irrespective of the sequence and the initial condition, and very close
to the measured value of $T_{1}$. This is a verification of the assumption
that all sequences can effectively decouple the slowly fluctuating environment.

For shorter pulse delays (i.e. more pulses in a given time interval), the
signal decays more rapidly. This is most prominent for the CP sequence.
In this situation, pulse imperfections add coherently and generate a rapid
loss of magnetization \citep{Alvarez2010}.

\begin{figure}[h]
\begin{centering}
\includegraphics[clip,width=0.9\columnwidth]{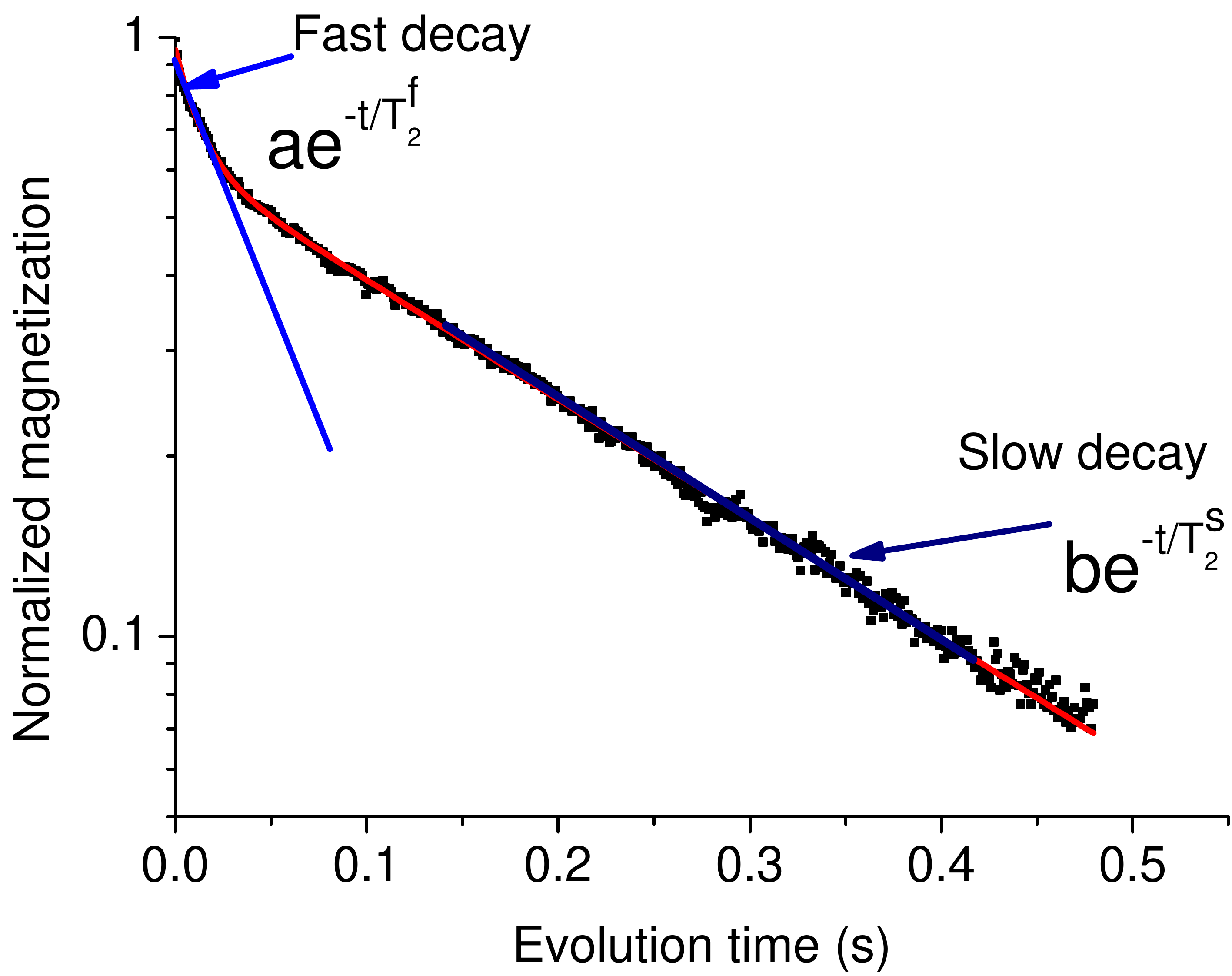}
\par\end{centering}

\caption{(Color online) Normalized spin magnetization as a function of the evolution
time for short delays ($\tau=100\,\mu$s) between the pulses for an XY8(s)
sequence. Pulse errors dominate here, inducing a multi-exponential decay\@.
The red solid line is a fit to \lyxadded{mustafa}{Sun Mar 10 14:01:39 2013}{E}q.~(\ref{eq:DoubleExp}).\label{5}}
\end{figure}

As the pulse delays become shorter than 0.5 ms, which corresponds to 864
pulses during the 0.5 s measurement time, the other sequences also start
to generate shorter decay times, and their decays become nonexponential.
Figure\ \ref{5} shows a representative example of such a signal. It can
be fitted with a double exponential,
\begin{equation}
s(t)=a\, e^{-t/T_{2}^{f}}+b\, e^{-t/T_{2}^{s}}\label{eq:DoubleExp}
\end{equation}
with two decay times $T_{2}^{f}$ and $T_{2}^{s}$.

\begin{figure}[h]
\begin{centering}
\includegraphics[width=1\columnwidth]{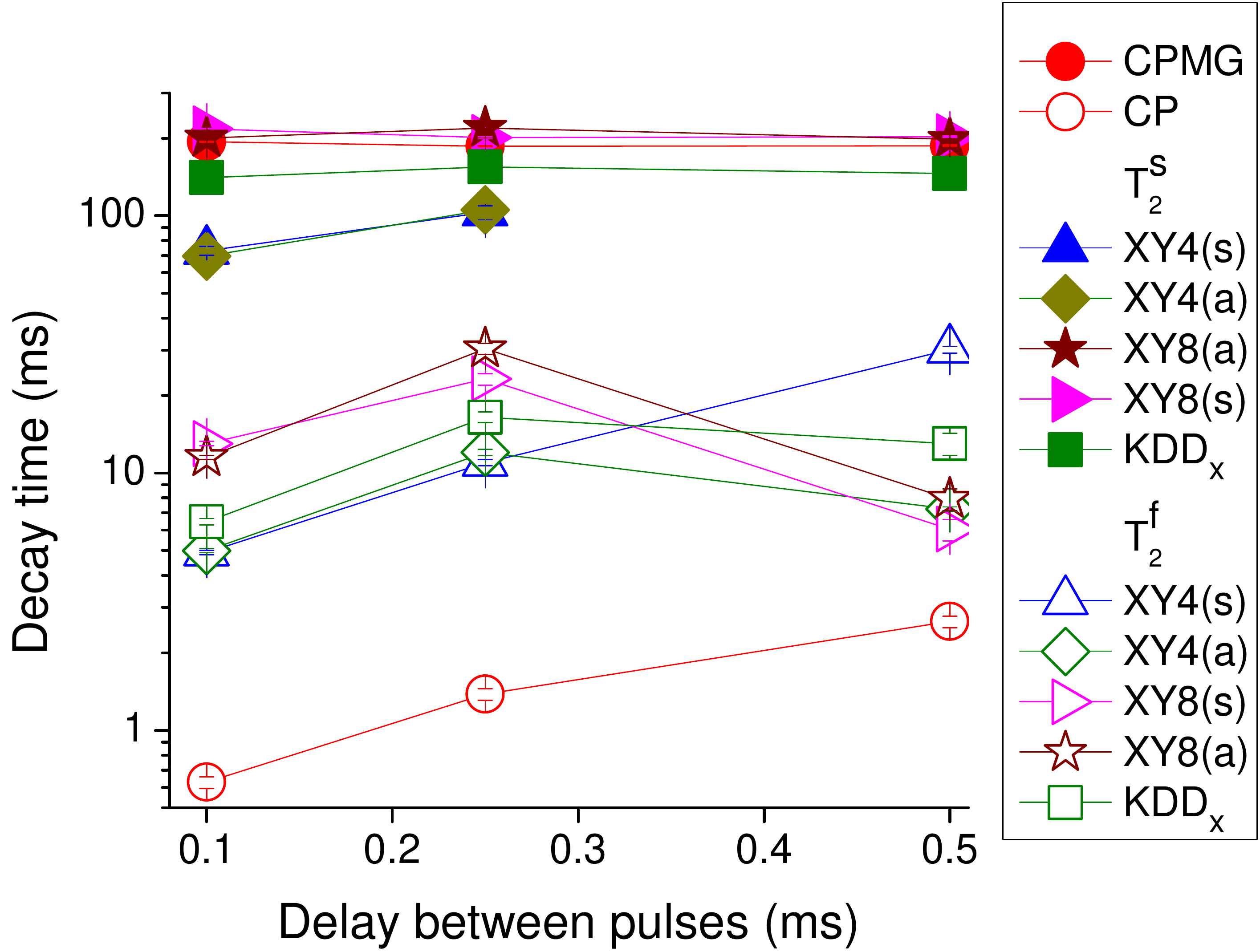}
\par\end{centering}

\caption{(Color online) Fitted decay times for the double-exponential decay\label{6}}
\end{figure}

Figure\ \ref{6} shows the decay times fitted with \lyxadded{mustafa}{Sun Mar 10 14:03:02 2013}{E}q.\ (\ref{eq:DoubleExp})
for different average delays between pulses\@. For the fast decay times
($T_{2}^{f}$), which are represented by empty symbols in Fig.\ \ref{6},
the performance of all DD sequences is quite similar\@. For the slow component
$(T_{2}^{s}$, represented by filled symbols), XY8, CPMG, and KDD$_{x}$
perform better than XY4.

As we stated before, the decay time is dominated by the effect of pulse
errors for short delays. The resulting average Hamiltonian projects the
magnetization onto its eigenbase - this results in the fast decay component.
After this projection, the remaining magnetization, which is not significantly
affected by the pulse imperfections, decays on a slower timescale, which
is dominated by the environment.

For a quantitative comparison of the different pulse sequences, we calculate
the average magnetization decay resulting per pulse of the sequence. For
this evaluation, we consider only the short-time component described by
$T_{2}^{f}$. 

Since the pulse error is the dominant source of decay, we quantify its
effect by measuring the fractional decay of the magnetization per pulse.
The pulses are the same for all the DD sequences, but their effect, averaged
over full cycles, shows how well the sequence is able to cancel the imperfections
of the individual pulses.

\begin{figure}[h]
\begin{centering}
\includegraphics[clip,width=0.95\columnwidth]{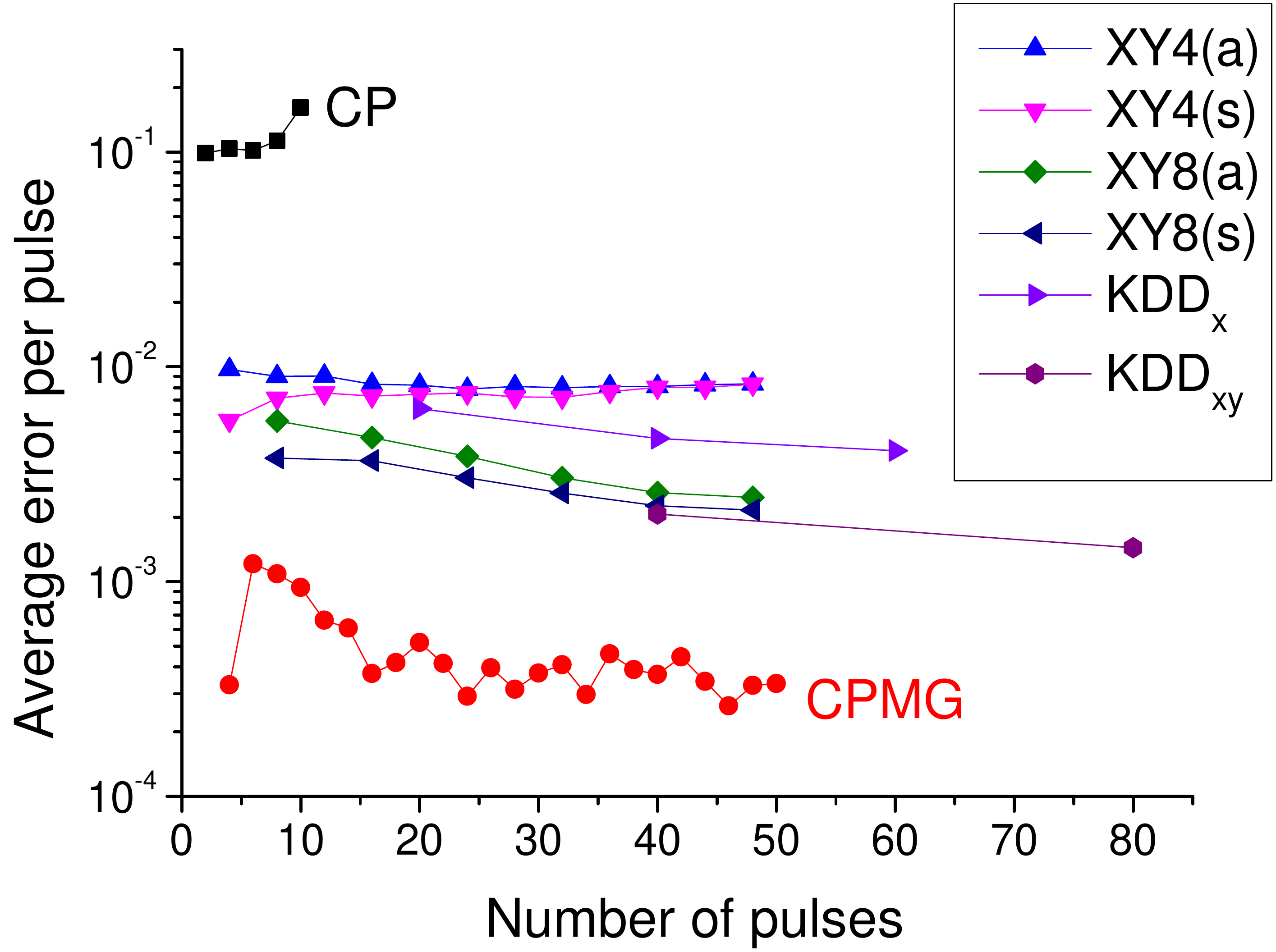}
\par\end{centering}

\caption{(Color online) Average error per pulse for different DD sequences with
delay $\tau=$100$\mu s$.\label{7}}
\end{figure}

Figure \ref{7} shows the average decay per pulse for the different sequences,
plotted against the number of pulses. For these data, the interpulse delay
was $\tau=$ 100 $\mu s$. The most conspicuous feature is that CP performs
very badly and CPMG very well. The compensated sequences lie between these
two extremes, and we find that the higher order sequences (XY8, KDD perform
better than the lower order sequences (XY4). For unknown initial conditions,
KDD shows the best performance. Under the present conditions, sequences
that differ only with respect to time reversal symmetry perform quite similarly,
in contrast to other cases discussed earlier \citep{Souza2012b}.

\section{Theoretical analysis by average Hamiltonian theory}

Average Hamiltonian theory (AHT) can be used to describe the effect of
DD sequences applied to the system during an interval of time $t_{1}<t<t_{2}$\@.
If the evolution of the system is governed by a time-dependent Hamiltonian
$\mathcal{H}(t)$, the effective evolution can be described by an average
Hamiltonian $\widetilde{\mathcal{H}}(t_{1},t_{2})$\@. If the Hamiltonian
$\mathcal{H}(t)$ is periodic with a cycle time $\tau_{c}$, i.e. $\mathcal{H}(t)=\mathcal{H}(t+\tau_{c})$
and the observation is stroboscopic and synchronized with the period $\tau_{c}$,
the evolution operator for each cycle is $\exp\left\{ -i\widetilde{\mathcal{H}}(0,\tau_{c})\,\tau_{c}\right\} $~\citep{Haeberlen1968,Ernst1987}.

As we discussed above, the remaining environmental noise fluctuates so
rapidly that its effects cannot be reduced by DD. It is a good approximation
to describe it as a classical field affecting the precession frequency
of the spins~\citep{Abragam1983}. Therefore Eq.~(\ref{eq:Hs}) can be
written as
\begin{equation}
\mathcal{H}_{SE}=\Delta\omega_{z}(t)S_{z},\label{eq:3}
\end{equation}
where the average of the random precession frequency is $\left\langle \Delta\omega_{z}(t)\right\rangle =0$
for every $t$. It causes an exponential attenuation $e^{-t/T_{2}}$ independent
of the delay between the pulses.

We can write the pulse propagator as a composition of the product of the
ideal pulse propagator $R_{\phi}=e^{-i\pi S_{\phi}}$ and two additional
evolutions for flip angle errors as

\begin{equation}
R_{\phi}=e^{-i(1+\epsilon)\pi S_{\phi}}=e^{-iH_{\phi}t_{p/2}}e^{-i\pi S_{\phi}}e^{-iH_{\phi}t_{p/2}},\label{eq:4}
\end{equation}
where $H_{\phi}=\frac{\epsilon\pi}{t_{p}}S_{\phi}$ and $t_{p}$ is the
pulse length\@. For the sequences XY4(s) and XY4(a) the effect of the
flip angle error vanishes in the zero-order average Hamiltonian, while
the first-order term for both sequences is~\citep{Souza2012b}
\begin{equation}
\widetilde{\mathcal{H}_{1}}^{XY4(s)}=\widetilde{\mathcal{H}_{1}}^{XY4(a)}=\frac{5\epsilon^{2}\pi^{2}}{16\tau}S_{z}.\label{eq:5}
\end{equation}
This shows that there is no difference between symmetric and asymmetric
sequences of XY4 up to first order of the average Hamiltonian, which is
in good agreement with the experimental results of Fig.~\ref{7}\@.

We consider now the XY8(s) and XY8(a) sequences. The zero-order and the
first-order average Hamiltonian terms in Ref.~\citep{Souza2012b} vanish
if we consider only the flip angle error effects\@. The first non-zero
term of the average Hamiltonian is then the second-order term, which is
again equal for both versions of the sequence:

\begin{equation}
\widetilde{\mathcal{H}_{2}}^{XY8(s)}=\widetilde{\mathcal{H}_{2}}^{XY8(a)}=\frac{13\epsilon^{3}\pi^{3}}{1536\tau}(S_{x}+S_{y}).\label{eq:6}
\end{equation}
This is also in excellent agreement with Fig.~\ref{7} where the symmetric
and asymmetric version behave similarly, but they are more robust than
the XY4 sequences. 

For CPMG and CP, the zeroth-order and first-order average Hamiltonians
are

\begin{equation}
\widetilde{\mathcal{H}_{0}}{}^{CPMG}=\frac{\epsilon\pi S_{y}}{\tau},\label{eq:7}
\end{equation}

and
\begin{equation}
\widetilde{\mathcal{H}_{1}}{}^{CPMG}=0.\label{eq:8}
\end{equation}

In the CPMG experiment, the initial condition is $\propto S_{y}$, which
commutes with the average Hamiltonian and is therefore not affected by
pulse errors. In the CP experiment, the initial condition is $\propto S_{x}$,
which is dephased by the pulse errors, in agreement with the data in Fig.~\ref{7}~\citep{Alvarez2010}. 

In the case of the KDD$_{x}$ pulse sequence, the zero- and first-order
average Hamiltonians vanish. The higher order terms were kept small by
the design of the sequence~\citep{tycko:2775,Jones2013}. As shown in
Ref.~\citep{Ichikawa2012}, this makes it robust against several systematic
errors because it is a geometric quantum gate. In Ref.~\citep{Souza2011},
we showed by numerical simulation and experimental data that the KDD$_{xy}$
sequence is more robust against flip angle errors than the other DD sequences
tested. Overall, the experimental comparison between the different sequences
is in good agreement with the numerical simulations and analytical results
based on average Hamiltonian theory\@.

\section{Conclusion}

We have tested the robustness of different DD sequences by comparing them
in an environment that interacts with the spins in such a way that the
decoherence time under the application of DD sequences with ideal pulses
is independent of the delay between the pulses. This allowed us to study
the robustness of the different DD sequences by isolating the effects of
the pulse errors. We found that the decoherence time of the most robust
sequences, the KDD family, is the longest for arbitrary initial states.
This is consistent with the measured error per pulse averaged over many
cycles of a DD sequence, where the KDD sequences have the lowest effective
error. In the regime studied, the time symmetrization on the cycles does
not play a significant role for reducing decoherence, since only the phases
of the pulses are important for reducing the effect of pulse errors. Our
experimental results for pulse errors are in good agreement with previous
numerical simulations and predictions of average Hamiltonian theory\@.

\emph{Acknowledgments: } We thank Alexandre M. Souza for helpful discussion\@.
This work was supported by the DFG through grant Su 192/24-1.\\

\bibliographystyle{apsrev4-1}
\bibliography{Ref}

\end{document}